\documentstyle[12pt]{article}
\newcounter{mycount}

\newcommand{\be}{\begin{eqnarray}}
\newcommand{\ee}{\end{eqnarray}}
\newcommand{\bfl}{\begin{flushleft}}
\newcommand{\efl}{\end{flushleft}}
\newcommand\ie {{\it i.e. }}
\newcommand\eg {{\it e.g. }}
\newcommand\etc{{\it etc. }}

\newcommand\half{\frac 1 2 }
\newcommand\noi{\noindent}

\begin{document}

\begin{flushright}
USITP 94-11 \\
July 1994
\end{flushright}
\centerline{\Large\bf The QCD Trace Anomaly as a Vacuum Effect}
\vskip 2mm
\centerline{\large\bf (The vacuum is a medium is the message!)}
\vskip 28mm
\centerline{\bf J. Grundberg$^\star$ and T. H. Hansson$^\dagger$}
\vskip 5mm
We use arguments taken from the electrodynamics of media to deduce the
QCD trace anomaly from the expression for the vacuum energy in the presence of
an external color magnetic field.
\vskip 8cm
\noi Submitted to: {\em American Journal of Physics}
\vfill
\noindent
$^\star$ Department of Mathematics and Physics, M{\"a}lardalens H{\"o}gskola
\\  \ Box 11, S-72103 V{\"a}ster{\aa}s, Sweden \\730  Sweden \\
$^{\dagger}$ Supported by the Swedish Natural Science Research Council
\eject

\noi
{\bf I. Introduction}

Intuitive discussions of phenomena in quantum field theory frequently take the
form of describing them as properties of the vacuum. The vacuum, it is
then argued, is just like any other medium, so we can use the intuition
developed in other fields of physics to get an understanding of quantum field
theory. Perhaps the first example of this was the interpretation of unoccupied
negative energy states in Dirac's relativistic electron theory as positrons.
Today the obvious analogy is the behavior of electrons in semiconductors,
though historically it seems that the concept of holes developed independently
in quantum field theory and solid state physics \cite{BAYM}.
Another example is the Casimir effect,
which is explained in terms of the zero-point fluctuations of the
quantized electromagnetic field. A third example is spontaneous symmetry
breaking of the vacuum, where the intuition is supplied by the behavior of
(anti-)ferromagnets and (for local symmetries) superconductors.
As a final example we have the charge-screening
properties of the vacuum. Already in the 30s it was realised that the QED
vacuum behaves just like an ordinary dielectric medium \cite{WEISS}:
Charges placed in the
vacuum get screened by polarization effects. Around 1980 Nielsen, \cite{NIELS},
and Hughes, \cite{HUGHES},
showed that asymptotic freedom, the antiscreening property of the QCD vacuum,
could be interpreted as (color-) paramagnetism.\footnote{Apparently
this had earlier been realised by 't Hooft, but not published
by him \cite{THOOFT}. }
In this paper we will
show that the calculation this result rests upon, the change in vacuum energy
due to an external (abelian color-) magnetic field, can also be used to
derive the QCD trace anomaly. The point is that the energy is all we need
in order to
apply some elementary arguments borrowed from the electrodynamics of media
to get the stress tensor, and hence the
anomaly. The vacuum is indeed just another (color-) electromagnetic medium.

Our paper is organized as follows. We first (section 2) review the
medium picture of screening (QED)  and antiscreening (QCD).
In the course of this discussion we quote the the result for the
vacuum energy in QCD and introduce the renormalization group equation for
charges. In section 3 we then use the expression for the vacuum energy and
some "elementary physics" arguments to
deduce the the stress tensor for QCD and calculate the anomaly. In these two
sections we have tried to keep the discussion at an elementary level
and to explain those results that we have just quoted from the literature.
Finally, in section 4, we
conclude with some comments as to how our derivation of the trace anomaly
reflects the breaking of scale invariance.
In appendix A we give a standard derivation of the trace anomaly and
compare some details with our derivation. We also briefly discuss higher
order effects and some other technicalities. In appendix B we justify an
argument based on Lorentz invariance used in the text. Both appendices
requires more knowledge of quantum field theory than the main text.

Although this work is a rederivation of a known result, we hope that it
has a value beyond the purely pedagogical.   The vacuum as a
medium has been a fruitful picture in trying to develop intuition about non
Abelian gauge theories.  In order to make good use of the medium picture,
and avoid pitfalls due to false analogies, we  think it is of value to
establish its limits as precisely as possible.

\vskip 3mm \noi
{\bf 2. Charge (anti)screening}

As a preliminary to the dicussion of the trace anomaly we wish to remind
the reader about the intuitive interpretation of charge (anti-)screening.
Starting with QED, we consider two  heavy particles of charge $e$,
separated by a distance $r$. Classically the potential energy of the pair is
given by
\be
E(r) ={\alpha \over r}\ \ \ \ \ ,
\ee
where $\alpha = e^{2}/4\pi$.
This result is modified in QED. To O($\alpha ^{2}$) the potential energy is
instead given by
\be
E(r)={1 \over\epsilon (r)}{\alpha\over r}\ \ \ \ \ ,
\ee
with \cite{QED114}
\be
{1\over\epsilon (r)}= 1 - \frac \alpha {3\pi} \ln \frac {\Omega^2} {m^2}
+{2\alpha\over 3\pi}\int_{m}^{\infty}du\,
e^{-2ru}\left[{2u^2 + m^2\over 2u^{2}}\right]{(u^2 - m^2)^{1/2}\over u^2}\
\ \ \ \ ,
\ee
where we have introduced an ultraviolet cut-off $\Omega$. The usual charge
renormalization in QED is obtained by taking
\be
\alpha_{ren} = \alpha\left(1 - \frac \alpha {3\pi} \ln \frac {\Omega^2}
{m^2}\right)
\ee
as the physical charge. This amounts to removing the cut-off dependent term
from equation (3), and in the limit $r\rightarrow\infty$, we get
\be
{1\over\epsilon (r)}\rightarrow 1+{\alpha\over 4\sqrt{\pi}(mr)^{3/2}}
e^{-2mr}.\label{epsilon}
\ee
The "effective charge" defined by
\be
E(r) = {\alpha_{eff}\over r}\ \ \ \ \ ,
\ee
depends on the distance. The intuitive explanation of
this is that the QED vacuum behaves like a polarizable medium. In Dirac's
original picture the vacuum has a homogeneous distribution of negative energy
electrons. When an external charge is introduced, the charge distribution
is modified. More negative charge accumulates around a positive external
charge, less around a negative external charge. The effect is that the
external charge is screened. This way of looking at the problem is not
explicitly charge-symmetric, but the same intuitive picture works if
we  prefer to view the vacuum a consisting of (virtual) electron-positron
pairs.

 The quantitative result for $\epsilon (r)$ given in (\ref{epsilon})
has the property that $\epsilon\rightarrow 1$ as $r\rightarrow\infty$.
This limit is finite
because the electron mass $m\not= 0$ and we have used this to
renormalize the electric charge in such a way that $\alpha$ means the charge
at the "preferred" distance $r=\infty$, but
this is clearly a matter of choice. A situation more like QCD arises
when $m=0$. In the limit $m\ll 1/r$, we get
\be
\delta E (r) = \frac \alpha r \left[- \frac \alpha {3\pi} \ln \frac
{\Omega^2}{m^2} + {2\alpha\over 3\pi }
\int_{m}^{\infty} du {e^{-2ru}\over u} \right]\ \ \ \ \ .
\ee
Note that both terms are logarithmically divergent as $m\rightarrow 0$, but
that the sum is finite. The charge renormalization can now be done in the
following way: First, note that in the limit $m\rightarrow 0$,
\be
r\delta E (r) -r_{0}\delta E (r_{0}) =
-{2\alpha^{2}\over 3\pi}\ln {r\over r_{0}}\ \ \ \ \ ,
\ee
so we can write  the total potential energy as,
\be
E  &=& {\alpha\over r} + \delta E (r)\cr\cr
&=&{\alpha +r_{0}\delta E(r_{0}) \over r}
-{2\alpha ^{2}\over 3\pi r}\ln {r\over r_{0}}\ \ \ \ \ .
\ee
Then we define
\be
\alpha (r_{0}) = \alpha + r_{0}\delta E(r_{0})\ \ \ \ \ ,
\ee
to get, to O($\alpha^{2}$),
\be
E(r)= {\alpha (r_{0}) \over r}\left(1 - {2\alpha (r_{0})\over 3\pi}
\ln{r\over r_{0}}\right)\ \ \ \ \ .
\ee
It is no longer possible to take the limit $r_{0}\rightarrow\infty$
in defining the charge, instead we pick an arbitrary scale $r_{0}$. The
dependence of the charge parameter
on the defining distance is governed by the statement
that the energy $E(r)$ is a physical quantity and should be independent of
arbitrary choices like $r_{0}$. Thus
\be
{dE\over dr_{0}} = 0\ \ \ \ \ ,
\ee
which, to this order in $\alpha$, gives
\be
r_{0}{d\alpha\over dr_{0}} = -{2\alpha^2\over 3\pi}\ \ \ \ \ .
\ee
This is the renormalization group equation for $\alpha$. It is conventional to
write renormalization group equations in terms of derivatives with respect to a
mass scale $\mu$ rather than a length scale. This merely reverses the sign
of the right hand side of the equation and we get
\be
\mu {d\alpha\over d\mu} = \beta (\alpha) = {2\alpha ^2\over 3\pi}\ \ \ \ \ .
\ee

The description of screening, as formulated above, is in terms of electric
fields
from a point charge. One can also discuss the effect in terms of the response
of the vacuum to an external homogeneous electromagnetic field. Since an
external electric field always gives a non-vanishing probability for pair
creation, it is more convenient to discuss the response to an external
magnetic field and calculate the magnetic permeability $\mu$ (not to be
confused with the mass scale in the renormalization group equation!)
rather than
$\epsilon$. We can then deduce $\epsilon$ from $\mu$ if we assume that the
vacuum is Lorentz-invariant. If that is the case
(for further discussion, see Appendix B), then we should have
\be
\epsilon\mu =1 \ \ \ \ \ ,
\ee
so, if $\mu>1$ (paramagnetic vacuum), then $\epsilon <1$ (antiscreening)
and if $\mu<1$ (diamagnetic vacuum), then $\epsilon>1$ (screening). The
strategy is thus to calculate the energy of the vacuum
in an external magnetic field and read off $\mu$. This was done in
\cite{NIELS}, \cite{HUGHES},
for the case of a constant homogeneous color magnetic field
belonging to an abelian subgroup of the color group, \ie of the type
\be
{\bf A}^a(x) &=& {\bf A}(x) T^a \\
{\bf B}^a(x) &=& {\bf\nabla}\times{\bf A}(x) T^a
\ee
where $T^a$ is a constant color matrix which is usually taken to be
diagonal (\ie for SU(2) proportional to $\sigma^3$
and for SU(2) \eg  proportional to  $\lambda^8$).
We refer the reader to the above  references for the details,
here we will just outline the calculation and give the final result.

The vacuum energy is calculated by summing the zero-point energies for the
modes of all fields. For the gluon field this is ${\hbar\omega\over 2}$ for
each mode, the usual harmonic oscillator zero-point energy,
for the quarks the exclusion principle (technically, the fact that fermion
creation/annihilation  operators satisfy anticommutaton relations instead
of commutation relations)
reverses the sign of this, so
the contribution is $-{\hbar\omega\over 2}$ for each mode, counting particles
and antiparticles separately.   In the Dirac sea picture, one can
also think
of the fermionic part as coming from the filled negative energy states,
where each mode contributes $-{\hbar\omega}$.

The frequencies for
the various modes are essentially the energy eigenvalues for particles with
spin 1 (gluons) or spin 1/2 (quarks) in an external magnetic field. As is well
known from elementary solid state physics (in the case of spin 1/2),
both the spin
and the orbital magnetic moment give contributions. Remember that without
spin, an ordinary electron
gas would be diamagnetic, with the spin included it is actually paramagnetic.
In the case of the vacuum medium the situation is reversed.
The vacuum electrons give a diamagnetic contribution due to the
extra sign coming from the exclusion principle. For
gluons the story is similar, but since there is no  extra minus sign, their
contribution is paramagnetic.
Of course, when we sum over all modes, the energy actually
diverges. One has to introduce a cut-off $\Omega$ and subtract a piece
$\propto \Omega^4$ that is independent of the external field. The result for
the field-dependent piece in QCD with $N_{F}$
species of mass-less colored fermions is
\be
U_{vac} = {V\over 2}\left[B^2 +(gB)^2 {33-2N_{F} \over 48\pi^2 }
\ln{\Omega^2\over |gB|}\right] \ \ \ \ \ ,
\label{energy}
\ee
where $g$ is the QCD coupling constant and V is the volume of the box whose
energy we have calculated. The first term in (\ref{energy}) is the classical
contribution to the energy,
just like the Coulomb contribution in the case of
external point charges. We still have an explicit $\Omega$ in this expression.
We can handle it in a way that is similar to what we did in the point charge
case above and which allows us to let $\Omega \rightarrow\infty$.
By rearranging
the expression in (\ref{energy}) we can write it as follows:
\be
u_{vac}={U_{vac}\over V}=&&{1\over 2} B^2\left[1 + g^2{33-2N_{F}\over 48\pi^2}
\ln{\Omega ^2\over |gB_{0}|}
\right]+\cr\cr &&{1\over2}(gB)^2{33-2N_{F}\over 48\pi^2}\ln|
{gB_{0}\over gB}|\ \ \ \ \ \ .
\ee
We now renormalize the field strength, in analogy with (10), by defining
\be
B_{ren}(gB_{0})
= B\left[1+g^2 {33-2N_{F}\over 96\pi^2}\ln{\Omega^2\over|gB_{0}|}
 \right] \ \ \ \ \ ,
\ee
so, to O($g^2$)\ \ \ \ \ ,
\be
u_{vac} = {B(gB_{0})^2\over 2}\left[1+g^2{33-2N_{F}\over 48\pi^2}
\ln|{gB(gB_{0})\over gB_{0}}|\right]\ \ \ \ \  ,
\ee
where all field strengths are renormalized ones.
Just as we did for the charge in QED, we can now
derive a renormalization group equation for $B(gB_{0})$. The result is
\be
gB_{0}{dB\over d(gB_{0})} = {33-2N_{F}\over 96\pi^2}g^2 \ \ \ \ \ .
\ee
If we remember that $gB$ is a physical quantity, we can readily
convert this to a renormalization group equation for g: $g_{r}B_{r}=gB$,
so
\be
gB_{0}{dg_r\over d(gB_{0})} =- {33-2N_{F}\over 96\pi^2}g_3^3
= {\beta(g_r)\over2}\ \ \ \ \ ,
\ee
where the coupling constants are renormalized ones.
We have introduced the conventional $\beta$-function, and the factor 1/2
in the last equality arises because $gB_{0}$ has dimension (mass)$^2$, and, as
we said above, it is conventional to use the derivative  with respect to
something with dimension mass.

To deduce a permeability from the vacuum energy, one usually defines $\mu$ by
\be
u = {1\over 2} HB ={1\over 2\mu} B^2\ \ \ \ \ ,
\ee
where the first equation has been taken from the electrodynamics
of linear media. This is strictly
speaking not quite correct, and we will return to the relation between $H$,
$B$ and $u$ in the next section, but for the present purposes it suffices.
We get
\be
\mu = 1 + {33-2N_{F}\over 48\pi^2} \ln|{gB_{0} \over gB}|\ \ \ \ \ ,
\ee
so we see that if $N_{F}<16$, $\mu>1$, the QCD vacuum is paramagnetic, and, by
the previous argument $\epsilon <1$, the QCD vacuum antiscreens color charge.

\vskip 3mm \noi
{\bf 3. The trace anomaly }

The point of view we wish to emphasize in this paper is that the QCD
vacuum is like any other nonlinear medium. In particular, once we have the
energy as a function of the appropriate variables, we can use elementary
virtural work arguments to deduce the stress tensor and thus calculate the
trace anomaly. However, before turning to that argument we shall ask what
the medium we are talking about really is.

In ordinary electrodynamics for media, the inhomogeneous
Maxwells equations are given in
terms of the  fields $\vec D$ and $\vec H$ and the corresponding
macroscopic currents and charges. The induced current and charge
distributions in the medium is taken into account via polarization and
magnetization which relates the $\vec D$ and $\vec H$ fields to the
fields $\vec E$ and $\vec B$. Since the   fields and
the macroscopic currents are related by Maxwells equations, we can choose
to either consider the currents or the fields as fixed. When discussing the
vacuum the situation is less clear, since there is no obvious separation
between external (or macroscopic) and internal (or microscopic) currents.
Roughly speaking, it is the virtual particles that constitute
the medium, but we have to make this statement a bit more precise.
In QED the situation is rather simple if we consider the case of an
external field and work only to lowest order in the coupling constant. In
this case the only virtual particles are electron - positron pairs, and the
calculation is exactly the one given in the previous section. Since there
are no virtual photons, there is no ambiguity in how to define the
background field. Note that if we had choosen to start from external
currents of electrons rather than from a background field, we would have
had to face the problem how to distinguish the ''real'' (on shell)
electrons in the currents from the virtual (off shell) electrons in the
vacuum medium.

In the case of QCD the situation is more subtle. Here
there are virtual gluons produced to lowest order, and these have to be
distinguished from the gluons that constitute the external field. This is
done by employing the so called background field method.
(For a review see \eg \cite{abbot}). This allows for
a separation of the gluon field in a background and a quantum part. Only
the quantum part appears in loops and needs to be gauge fixed.
The background (or external) field is treated as classical, but
due to the quantum loops it gets a non-quadratic (but gauge-invariant)
effective action. The separation between background and quantum part of the
gluon field depends on the gauge-fixing and renormalization procedure and
is thus to a certain extent arbitrary. Nevertheless it will provide a strict
definition of macroscopic  and microscopic and thus allows us to use
results from the electrodynamics of media, just as in the case of QED.

In order to identify the background field, which was introduced in (17), and
that appears in the expression (21) for the vacuum energy, we first note that
the field (17) which enters our expressions is the curl of the vector
potential (16). Thus it is
divergence free, and should be  a ${\bf B}$ rather than a {\bf H}-field.
Second, our complete hamiltonian density, including the background field
energy, is
(for the time being we put a tilde on the field to remind us that our aim
is to identify it)
\be
{\cal H} ={{\tilde B} ^{2}\over 2} +{\cal H}_{matter}
({\bf p}-g{\bf A}_{quant}-g{\bf {\tilde A}})\ \ \ \ \ ,
\ee
where ${\cal H}_{matter}$ is the hamiltonian density for quarks
and (the quadratic part of the) hamiltonian for gluons in an
external magnetic field.
For a constant and homogeneous external field we can choose the vector
potential ${\bf\tilde A} = \half {\bf {\tilde B}\times r}$.
Varying the field we get
\be
\delta u_{vac} = \delta \langle vac |{\cal H}  | vac \rangle =
({\bf {\tilde B} } -{\bf M} )\cdot \delta {\bf \tilde B }\ \ \ \ \ ,
\ee
where we have identified the magnetization,
\be
{\bf M} = \langle vac |-\half\, {\bf r} \times
\frac  { \partial{\cal H}_{matter} }
{\partial {\bf\tilde A}}  | vac \rangle  =
\langle vac |\half\, {\bf r} \times {\bf j} | vac \rangle \ \ \ \ \ ,
\ee
where $\bf j $ is the current operator. In the
non relativistic case the right hand side in (28)
reduces to a sum of the well know
expressions for  the orbital and and spin magnetic moment densities.
In the relativistic case, relevant for light quarks, there is no clean
separation between orbital an spin contributions on the operator level, but
the distinction can still be made by considering the following
expression for the (unrenormalized) one loop vaccum energy\cite{NIELS}
\be
u_{vac} =\half B^2 -\half B^2 (-1)^{2a}\frac {\hat g^2}
{4\pi^2} \sum_{m=-s}^s \left(
 m^2 - \frac 1 {12} \right) \ln |\frac {\Omega^2} {gB}| \ \ \ \ \  ,
\ee
which is valid for spin $s=0,\half,1$, and where $\hat g$
depends on the color representation of the particles. The
term $\frac 1 {12}$ comes from the orbital motion and is common for
both scalar, spinor and vector particles, while the term $m^2$, where $m$
is the component of the spin in the direction of the $\bf B$ field,
does depend on the spin. For gluons this expression coincides with (18)
if $\hat g^2 =\frac 3 2 g^2 $, corresponding to adjoint gluons.

To identify ${\bf H}$ and ${\bf B}$ we compare $\delta u_{vac}$
with the work done as the
magnetic field is varied. As the field changes an electric field ${\bf E}$
is induced and the work done on the system is \cite{JACKSON},
\be
\delta W &=& -\delta t \int d^{3}x\, {\bf E }\cdot{\bf j} =
\int d^{3}x\, \delta {\bf A }\cdot{\bf j }  \cr\cr &=&
\int d^{3}x\, \delta{\bf A}\cdot (\nabla \times {\bf H})   =
\int d^{3}x\,  {\bf H}\cdot\delta {\bf B} \ \ \ \ \ ,
\ee
where we have used Maxwells equation
\be
\nabla\times{\bf H} = \frac 1 c\,  {\bf j}\ \ \ \ \ ,
\ee
and ${\bf j}$ is the current density that serves as external source for the
magnetic field. The work per unit volume (30) should equal the change in
energy (27):
\be
{\bf H}\cdot\delta{\bf B} = ({\bf{\tilde B}} -
{\bf M})\cdot\delta{\bf{\tilde B}}\ \ \ \ \ .
\ee
Since ${\bf B}={\bf H}+{\bf M}$, we identify
\be
{\bf B} = {\bf{\tilde B}}\ \ \ \ \ ,
\ee
and
\be
{\bf H} = {\partial u(B)\over \partial {\bf B}}\ \ \ \ \ .  \label{H}
\ee
The work we discussed above is the total work on both medium and sources of
the magnetic field. To be able to deduce the stress tensor we need to consider
a situation where we can divide this work into work done on the medium and on
the sources. This is what we are going to do next.

In order to discuss the work done in changing the magnetic field it is
convenient to have a specific geometry in mind. We consider a cylindrical
solenoid of length $L$, radius $R$ and with $N$ turns. A current $I$ through
the solenoid generates a
${\bf H}$-field, which, in the limit $L\rightarrow\infty$,
$N/L$ finite, is homogeneous, parallell to the
axis of the solenoid and has magnitude
\be
H = \frac 1 c \frac N L I\ \ \ \ \ .
\ee
This solenoid encloses our medium, {\em i.e.}, the vacuum.
We can define the stress
tensor in terms of the virtual work done as we change the enclosed region while
keeping the current constant. (The more familiar electrostatic
analogue of this procedure is to consider a region bounded by capacitor plates
and the work done as the plates are moved while keeping them at constant
potential \cite{LANDAU}.)
First, it is clear from the symmetry of the problem that the only
non-vanishing components of the stress-tensor  $T_{ij}$ are $T_{zz}$ (where
the z-axis is the axis of the solenoid) and $T_{xx} = T_{yy}$. Thus $T_{ij}$
simply acts like an anisotropic pressure. As we change the solenoid we have to
do a virtual work per unit volume
\be
\delta w_{1} = "-p{\delta V\over V}"=
-T_{zz}{\delta L\over L} - T_{xx}{\delta (\pi R^{2})\over
\pi R^{2}}\ \ \ \ \ .
\ee
In addition to this work on the medium we also have to do some work
to keep the current constant. As the solenoid is modified the magnetic flux
changes and an emf,
\be
{\cal E} = - \frac{1}{c} {d\over dt} (B\pi R^2 ) \ \ \ \ \ ,
\ee
is induced. This emf does some work which has to be compensated for to keep
the current constant. The additional work per unit volume is given by
(\cite{JACKSON}, p 214)
\be
\delta w_{2} &=& - {NI \over \pi R^{2} L} {\cal E} \delta t =
{1\over\pi R^{2} Lc}{d\over dt}(\pi R^{2}NIB)\delta t\cr\cr &=&
{1\over \pi R^{2}L} \delta (\pi R^{2}LHB) \ \ \ \ \ ,
\ee
where we have used that $I$ is constant and the relation between $I$ and $H$.
The total work per unit volume is thus
\be
\delta w = \delta w_{1} + \delta w_{2} &=&
(HB-T_{zz}){\delta L\over L} + (HB-T_{xx}){2\delta R\over R}
+\delta (HB) \cr\cr &=&
(-H^{2}{dB\over dH} - T_{zz}){\delta L\over L} +
(HB-T_{xx}){2\delta R\over R}\ \ \ \ \ ,
\ee
where we have used that $H$, and consequently $B$, is independent of $R$ and
also that $H\propto L^{-1}$, \ie $L\delta H = -H\delta L$).
This work should equal the change in energy of the
system per unit volume. This change has two terms.
The first is due to that the
solenoid encloses a different volume, the second is due to the changed
magnetic field. The result is
\be
\delta w = u{\delta V\over V} + \delta u =
u{2\delta R\over R} + (u-H^{2}{dB\over dH}){\delta L\over L}\ \ \ \ \ .
\ee
In the last step we used that $\delta u = H\delta B$, \ie, equation (\ref{H}),
and again that $H\propto L^{-1}$.
We can conclude that,
\be
T_{zz} &=& -u\cr
T_{xx} &=& T_{yy} = HB-u\ \ \ \ \ ,
\ee
and hence the trace of the stress-energy tensor is
\be
T^{\mu}\thinspace_{\mu} = T_{00} - T_{xx} - T_{yy} - T_{zz} =
4u - 2HB\ \ \ \ \ .
\ee
{}From section 2 we know that
\be
u={B^{2}\over 2} + {33-2N_{F}\over 48\pi^2}
 { g^{2} B^{2} \over 2} \ln |{gB\over gB_{0}}|\ \ \ \ \ ,
\ee
and thus the final result
\be
T^{\mu}\thinspace_{\mu} = -{33-2N_{F}\over48\pi^2}g^2B^2=
{\beta (g)\over 2g }F^{\mu\nu}F_{\mu\nu}\ \ \ \ \ ,
\ee
where we have used the lowest order result for the $\beta$-function,
\be
\beta (g) = - {33-2N_{F}\over 48\pi^2} g^{3}\ \ \ \ \ .
\ee
We have thus recovered the usual result for the trace anomaly. Note that it
was important in this derivation to have the correct relation between $u$,
$B$ and $H$. If we had applied linear-medium relations, the anomaly would have
vanished. The entire effect comes from the non-linearity of the vacuum.

Readers wishing
for a comparison of the $T_{ij}$ we derived with a conventional field theory
version are referred to Appendix A, where we also show that the coefficient
in front of $F^{\mu\nu}F_{\mu\nu}$ is indeed related to the $\beta$-function.

\vskip 3mm \noi
{\bf 4. The trace anomaly and scale braking}

In the standard derivation of the trace anomaly, the relation to
spontaneously broken scale invariance is emphazised. At a classical level
the fields scale according to their canonical dimension,
\be
{\bf B}(x) \rightarrow \lambda^{-2} {\bf B}(\lambda x) \\
{\bf E}(x) \rightarrow \lambda^{-2} {\bf E}(\lambda x)
\ee
under the rescaling $x^\mu \rightarrow \lambda x^\mu$ of the coordinates.
Since the classical Lagrangian for the electromagnetic or gluon field does
not contain any dimensionful parameter, this scale transformation is a
symmetry and there is a corresponding conserved current, the so called
dilatation (or scaling) current given by
\be
j^\mu_{dil.} = T^\mu_\nu x^\nu \ \ \ \ \ .
\ee
Since the energy-momentum tensor is conserved, we get
\be
\partial_\mu j^\mu_{dil.} = T^\mu_\mu \ \ \ \ \ ,
\ee
so if the dilatation current is conserved, the trace of the energy-momenum
tensor vanishes. (For a more careful discussion of this statement, see
\cite{polc}.) The trace anomaly thus implies that the dilatation current is
not conserved. The classical scale invariance has been broken by quantum
effects. How did this happen? A look at our calculations shows that we at
some stage had to introduce a scale to be able to define the theory. In the
case of external charges in QED we needed the lengh scale $r_0$ to
define the charge, and in the case of an external homogeneous
magnetostatic field we introduced the scale in the guise of a reference
magnetic field $gB_0$. In both cases we made the point that this reference
scale ($r_0$ or $gB_0$) was arbitrary and we used this fact to derive
renormalization group equations.
However, despite the arbitrariness of the scale, it {\em had} to be introduced,
and its presence destroys the scale invariance.
As we change reference scale
and coupling constant, there is a combination of them that remains
unchanged: A renormalization group invariant scale that can be used to
parametrize the theory. By integrating the renormalization group
equation
\be
\mu \frac {dg} {d\mu} = \beta(g)\ \ \ \ ,
\ee
we can trade the arbitrary scale $\mu$ for a (dimensionful) integration
constant $\Lambda$
\be
\Lambda =  \mu e^{-\int_{g(\Lambda)}^{g(\mu)}
\frac {dg} {\beta(g)} }\ \ \ \  .
\ee
One might ask what is the advantage of having replaced the scale $\mu$ by
the (seemingly) equally arbitrary scale $\Lambda$. This becomes clear if we
reexpress the vacuum energy $u_{vac}$ in (21) in terms of a
$\Lambda_{B}$ defined via (51) with $\mu^2 = gB_0$:
\be
u_{vac} = -{33-2N_{F}\over 192\pi^2}   \Lambda_B^4
\ee
We see that $\Lambda_B$ is directly related to a physical quantity, while
$\mu$ is not. In fact, equation (51)
describes a a curve in the $g-\mu$ plane, and all points on this curve
corresponds to the same physics (masses, scattering amplitudes \etc). To
get different physics we have to change to a different curve, \ie change
the value of $\Lambda$. It is, however, important to realize that the
actual numerical value (in \eg MeV) of $\Lambda$ depends on how the
renormalization is performed. That is why we used a subscript $B$ on the
$\Lambda$ in (52). There are many different prescriptions for defining
$\Lambda$, but they can all be related, even though the required
calculations sometimes  can be lenghty.\footnote{
For instance, to relate the
$\Lambda$ that occurs naturally in lattice calculations to the one used in
perturbative QCD, one must perform a rather complicated
perturbative lattice calculation. }

Classically the theory is scale invariant and is characterized by a
dimensionless parameter, $g$. Quantum mechanics breaks the scale invariance
and transmutes $g$ to a dimensionful parameter, $\Lambda$! In QCD with mass
less quarks this is the only parameter that sets the scale for hadron
masses, crossections, {\em etc.}, and it is used to parametrize theoretical
predictions for QCD processes.

To summarize: The trace anomaly implies that scale invariance
is broken by quantum corrections, and this occurs because, in one way or
another, the renormalization procedure forces us to introduce a scale to
define the theory.

\vskip 3mm \noi
{\bf Acknowledgement}

We thank R.J. Hughes for a conversation about this and other possible ways
of giving an elementary discussion of the trace anomaly, and Ken Johnson
for helpful comments about the manuscript.

\vskip 15 mm \noi
{\large\bf Appendix A. }
\renewcommand{\theequation}{A.\arabic{equation}}
\setcounter{equation}{0}

In section 3 of this paper we used elementary arguments to deduce the
stress-tensor from the energy density. In this appendix we first show that
this way of deducing the stress-tensor coincides with the conventional field
theory definition in terms of a variation of the effective action with respect
to the metric. By combining the effective action method with renormalization
group arguments we then show that the coefficient in the trace anomaly is
indeed related to the $\beta$-function. (The argument here is the
inverse of the one presented in \cite{PAGELS}. We thank R.J. Hughes
for informing us about this reference.)

The starting point for our field theory considerations is the renormalized
effective action
\be
\Gamma = \int d^{4}x \sqrt{-g} \cal{L} \ \ \ \ \ .
\ee
The stress-energy tensor is then given by\footnote {
Note that the introduction of a
metric tensor is just a trick to calculate the symmetric energy-momentum
tensor. With some care, the same result can be obtained by canonical
methods. }
\be
T_{\alpha\beta}  = {2\over \sqrt{-g}} {\delta \Gamma\over \delta
g^{\alpha\beta}} =
2{\partial {\cal L} \over \partial g^{\alpha\beta}} -
g_{\alpha\beta}{\cal L} \ \ \ \ \ .
\ee
We are interested in $T_{\alpha\beta}$ for a constant (abelian color)
electromagnetic field, in which case ${\cal L}={\cal L}(x,y)$ where
$x=F^{\alpha\beta}F_{\alpha\beta}$ and $y=\epsilon ^{\alpha\beta\gamma\delta}
F_{\alpha\beta}F_{\gamma\delta}/\sqrt{-g}$ are the two possible invariant
combinations of field strengths. We get
\be
T_{\alpha \beta} =
4F_{\alpha} \thinspace^{\gamma} F_{\beta \gamma}
{\partial {\cal L} \over \partial x}
+ g_{\alpha \beta} \thinspace y {\partial {\cal L} \over \partial y }
- g_{\alpha \beta }{\cal L}\ \ \ \ \ .
\ee
In the following we will ignore the dependence on $y$ which is irrelevant
for the discussion in this appendix. In a purely magnetic field
\be
F_{0i} &=& 0\cr
F_{ij} &=& -\epsilon_{ijk} B^{k}
\ee
and thus
\be
T_{00} &=& -{\cal L}\cr
T_{0i} &=& 0\cr
T_{ij} &=& \delta _{ij} ({\cal L} - B^{k}{\partial {\cal L} \over \partial
B^{k}})
+{\partial {\cal L} \over \partial B_{i}}B_{j}\ \ \ \ \ ,
\ee
which we can compare with our "elementary physics" result from section 3
(slightly generalized),
\be
T_{00}&=& u\cr
T_{ij} &=& \delta _{ij} (-u + B^{k}{\partial u \over \partial B^{k}})
-{\partial u \over \partial B^{i}}B_{j}.
\ee
Thus the results do agree if $u=-\cal{L}$, as indeed it is \cite{WEISS}.
We can now
proceed to use the field theory calculation to show that the coefficient in
the trace anomaly really is the $\beta$-function. To this end it is convenient
to rescale the gauge fields by $gA_{\mu}\rightarrow A_{\mu}$. Then the
coupling constant only occurs as a coefficient in front of the gauge theory
action,
\be
S =
\int d^{4}x \{ - {1 \over 4g^{2}} F^{\mu\nu} F_{\mu\nu}+ ... \}\ \ \ \ \ ,
\ee
and we can define an effective coupling constant in the renormalized effective
action by writing
\be
\Gamma = \int d^4x\, {\cal L} =
   \int d^4x\,  -{1 \over 4g^{2}_{eff}} F^{\mu \nu}F_{\mu \nu}\ \ \ \ \ .
\ee
This expression defines an effective (or running) coupling constant
$g_{eff}$ which is a function of the original renormalized
coupling $g(\mu ),\mu$ and the invariants $x$ and $y$. Recall (section 2) that
$\mu$ is the renormalization scale. Since according to (A.5) $-\cal L$ is
nothing but the renormalized energy density of the vacuum, it is an
observable and thus cannot depend on the renormalization point $\mu$.

Combining our equations we get
\be
T^{\alpha}\thinspace _{\alpha} =
\left[ 4{x\over g_{eff}}{\partial\ g_{eff} \over \partial x}\right]
{1\over 2g_{eff}^{2}}x\ \ \ \ \ .
\ee
Dimensional analysis implies that the $\mu$-dependence of $g_{eff}$ is of the
form
\be
g_{eff}= g_{eff}({x\over \mu ^{4}},{y\over \mu ^{4}}, g(\mu))\ \ \ \ \ ,
\ee
and since $\cal L$ and thus  $g_{eff}$ does  not depend on the
arbitrary scale $\mu$, we get
\be
0 = \mu{dg_{eff}\over d\mu} =
-4x{\partial g_{eff}\over \partial x}
+\mu {\partial g \over \partial \mu}
{\partial g_{eff} \over \partial g}\ \ \ \ \ ,
\ee
and thus
\be
T^{\alpha}\thinspace_{\alpha} =
\mu {\partial g \over \partial \mu}{\partial g_{eff} \over \partial g}
{1 \over 2g_{eff}^{3}} F^{\mu\nu}F_{\mu\nu}\ \ \ \ \ .
\ee
The $\beta$-function is actually the component of a vector field in the space
of coupling constants:
\be
\beta (g) {\partial \over \partial g} =
\mu{\partial g \over \partial \mu }{\partial \over \partial g}\ \ \ \ \ .
\ee
It follows that
\be
\mu{\partial g \over \partial \mu}{\partial g_{eff} \over \partial g}
= \beta (g_{eff})\ \ \ \ \ ,
\ee
and thus
\be
T^{\alpha}\thinspace_{\alpha}=
{\beta (g_{eff}) \over 2g_{eff}^{3}} F^{\mu \nu}F_{\mu \nu}\ \ \ \ \ ,
\ee
which, up to the rescaling of the fields and to lowest non-trivial order in
perturbation theory, agrees with the result derived in section 3 and moreover
makes the relation to the $\beta$-function manifest.

The vacuum energy derivation of the trace anomaly given in section 3 makes
use of an expression for the energy that is essentially a one-loop result,
whereas the arguments in this appendix seems to involve no such
approximation. In principle we could calculate the background field
effective action to any desired loop order, and it can still always be
written in the form (A.8). For this conclusion we have to keep in mind that
there are two ways of presenting the anomaly. One is to establish an
equality between the operator $T^\mu_\mu$ and the renormalized field
operators of the theory. The other is to deal directly with matrix elements
for $T^\mu_\mu$ for states where there is an external gauge field. Both the
vacuum-energy derivation and the background field effective action
derivation follow the second way. It is known that as an operator equality
(A.15) is not true in general (beyond one loop). For further discussion, see
the review \cite{SHIFF} by M. A. Shifman.

\vskip 15 mm \noi
{\large\bf Appendix B. }
\renewcommand{\theequation}{B.\arabic{equation}}
\setcounter{equation}{0}

In section 2 we appealed to Lorentz invariance to get a relation between
$\epsilon$ and $\mu$. Actually it is not quite obvious that we can do so.
After all, we want the relation in the presence of an external field so the
physical situation is not Lorentz-invariant. We can use the effective action
formalism to see under what circumstances we can still use the naive argument.
It suffices for our purposes to consider the case of a slowly varying
external abelian color electromagnetic field. Then again ${\cal L}= {\cal
L}(x,y^2)$  with $x=2(B^2 -E^2)$ and $y=4 {\bf E}\cdot{\bf B}$. (Since $y$
is a pseudoscalar it can't occur in odd powers.) We get \cite{QED130}
\be
{\bf D} &=& {\partial {\cal L}\over\partial {\bf E}} =
-4{\partial {\cal L}\over\partial x}{\bf E}
+8y{\partial {\cal L}\over \partial y^2}{\bf B},\cr\cr
{\bf H} &=& -{\partial {\cal L}\over \partial {\bf B}} =
-4{\partial {\cal L}\over \partial x}{\bf B}
-8y{\partial {\cal L}\over \partial y^2}{\bf E}\ \ \ \ \ .
\ee
We see that if ${\bf E}\cdot{\bf B}=0$ and if we define
\be
{\bf D} = \epsilon {\bf E},\qquad
{\bf B} = \mu {\bf H}\ \ \ \ \ ,
\ee
we do get
\be
\epsilon\mu =1\ \ \ \ \ ,
\ee
(assuming the derivatives do not become singular as $y\rightarrow 0$).
In particular, we can use this relation for a homogeneous
magnetostatic field.

\end{document}